# A near infrared frequency comb for Y+J band astronomical spectroscopy

Steve Osterman[*a], Gabriel G. Ycas[b,c], Scott A. Diddams[b], Franklyn Quinlan[b], Suvrath Mahadevan[d,e], Lawrence Ramsey[d,e], Chad F. Bender[d,e], Ryan Terrien[d,e], Brandon Botzer[d,e], Steinn Sigurddson[d,e] and Stephen L. Redman[f]

[a] Center for Astrophysics and Space Astronomy, University of Colorado, Boulder, CO, USA;
[b] Time and Frequency Division, National Institute of Standards and Technology, Boulder, CO, USA;
[c] Department of Physics, University of Colorado, Boulder, CO, USA; [d] The Department of Astronomy, Pennsylvania State University, University Park PA, USA  [e] Center for Exoplanets and Habitable Worlds, Pennsylvania State University, University Park, PA, USA;  [f] Atomic Physics Division, National Institute of Standards and Technology, Gaithersburg, MD, USA

## ABSTRACT

Radial velocity (RV) surveys supported by high precision wavelength references (notably ThAr lamps and I2 cells) have successfully identified hundreds of exoplanets; however, as the search for exoplanets moves to cooler, lower mass stars, the optimum wave band for observation for these objects moves into the near infrared (NIR) and new wavelength standards are required. To address this need we are following up our successful deployment of an H band(1.45-1.7μm) laser frequency comb based wavelength reference with a comb working in the Y and J bands (0.98-1.3μm). This comb will be optimized for use with a 50,000 resolution NIR spectrograph such as the Penn State Habitable Zone Planet Finder. We present design and performance details of the current Y+J band comb.

## 1. INTRODUCTION

The National Academy of Science's Astro2010 decadal survey ranked the search for nearby habitable planets as one of the top three scientific objectives for the astronomical community and recommends 'aggressive development of ground-based high precision RV surveys of nearby stars at optical and NIR wavelengths.'[1]

The use of the radial velocity method for finding planets around main sequence stars could not have proceeded without the simultaneous development of high stability, high resolution spectrographs and of high precision wavelength calibration sources such as the thorium argon (Th-Ar) lamp and the iodine (I2) cell.[2,3] As the search for extrasolar planets moves to lower mass, cooler stars, new instruments and new technologies are needed to provide the requisite level of precision.  In response to this requirement we are developing a NIR laser frequency comb (LFC) optimized to support astronomical spectroscopy.

The overwhelming majority of planets identified to date orbit class F-K stars, with no more than 5% of the planets listed on the Exoplanet Data Explorer site orbiting cooler class M dwarf stars.[4] In light of this, there are compelling reasons to extend radial velocity planet searches to the NIR, particularly where the goal is the identification of a near-earth mass planet in the liquid water habitable zone:  First, M dwarfs are lower mass than other main sequence classes, so an earth mass planet at a fixed distance will induce a larger reflex motion in an M star than in a higher mass F-K class star. Second, since M dwarfs are cooler than F-K class stars, the habitable zone is closer to the star, leading to an increase in the orbital velocity of the planet and a corresponding increase in the stellar reflex velocity (fig. 1, left). Finally, M stars represent the majority of nearby stars (fig. 1, right).

---

[*] steven.osterman@colorado.edu; CASA, 1255 38th St, Boulder CO 80305; phone 1 303-492-3656

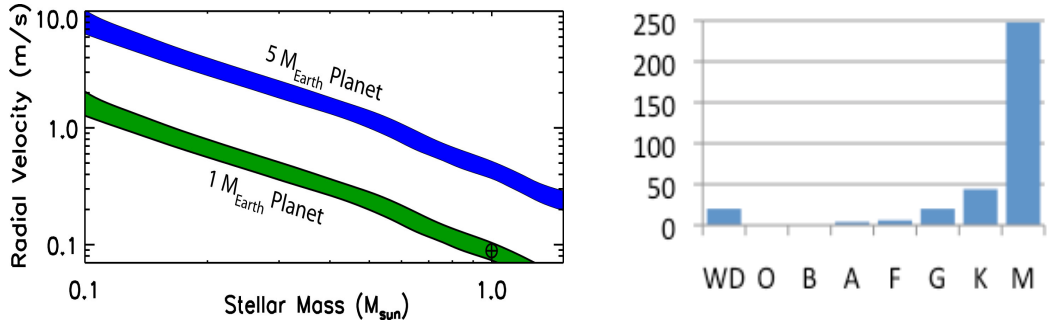

Figure 1: (left) Semi-amplitude of the Doppler wobble induced on the host star by a one and five earth mass planet in the habitable zone for an edge-on orbit (after Kasting, 1993.[5]). (right) Distribution of stars within 10 parsecs by spectroscopic class.[6]

M dwarf stars are cooler than the more studied F-K class stars, and as a result they are much dimmer in the visible than a hotter star at a comparable distance, and much dimmer in the visible relative to the infrared, requiring spectrographs optimized for the NIR. A spectrograph specifically designed to address this problem is currently under development at the Pennsylvania State University, the Habitable-zone Planet Finder (HPF).[7,8]

In order to take full advantage of the increased resolution and stability offered by the next generation of high precision NIR spectrographs, improved wavelength calibration sources are required. Wavelength references currently in use or in development in the NIR include absorption cells[9,10] and emission line lamps.[11] In addition, stabilized fiber optic Fabry–Pérot (FFP) filters are under investigation as a stand-alone wavelength reference.[12,13]

Absorption cells significantly complicate the stellar spectrum and data reduction pipeline, and at present none have been demonstrated for the Y and J band. Emission lamps are well suited for use as an injected wavelength standard for simultaneous calibrations in fiber fed instruments,[14,15] with U-Ne lamps[11] looking particularly promising. However, emission line lamps suffer from a number of limitations including irregularly spaced emission lines, widely ranging line intensities, potential age dependent effects and limited absolute wavelength knowledge.[16,17] The final wavelength standard mentioned above, stabilized FFP filters can provide uniformly spaced reference lines of roughly uniform intensity and tailorable line-to-line spacing. However, the FFP is much more strongly dependent on the stability of the filter cavity than are filtered LFCs since detuning a cavity from a series of discrete, narrow lines (as is the case with a frequency comb) induces less apparent RV error than is the case where the cavity is illuminated by a continuum source (fig. 2).

In contrast to these calibration sources, the LFC[18,19,20] has been show to provide a nearly ideal wavelength standard, with regularly spaced, bright laser emission lines spanning broad wavelength ranges and which is traceable to fundamental constants. This provides a calibration source that can provide a reproducible absolute frequency reference, with all LFCs traceable to the SI standard second. The frequency of the nth mode of a LFC is given by the relationship

$$f_n = n \times f_{rep} + f_{ceo},$$

where $f_{rep}$ is the mode-locked laser's repetition rate or mode spacing and $f_{ceo}$ is the carrier envelope offset frequency.[21] The precision of this relationship is limited only by the precision and stability of the RF reference used to control $f_{rep}$ and $f_{ceo}$. Field suitable references such as an Rb clock or a GPS disciplined oscillator can deliver frequency precision on the order of 1:10$^{-11}$, a level of accuracy that corresponds to sub-cm/s RV offset.[22] The line spacing can be tailored through the use of high finesse Fabry–Pérot (FP) cavities to provide easily resolved, distinct calibration lines, and bandwidth can to a large degree be expanded to meet the spectrograph's requirements, yielding a calibration source with precision and long term stability that significantly exceed the ultimate precision of the spectrograph (for example, $\Delta\lambda/\lambda < 10^{-11}$ for 10 cm/s RV determination).[23,24,25,26]

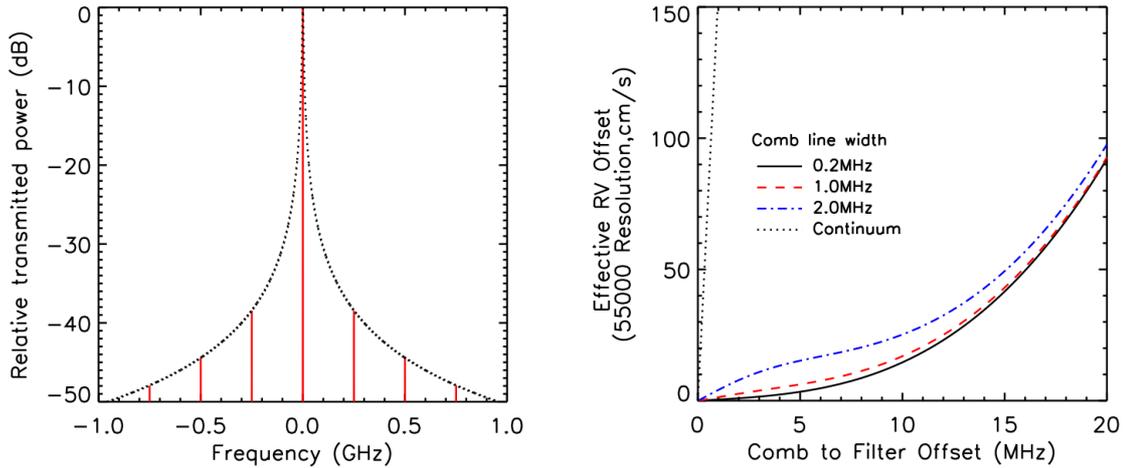

Figure 2: (left) Simulated Fabry-Pérot envelope (2000 finesse) applied to a 250 MHz LFC with 200 kHz wide lines at 1050 nm (typical of the broadened Yb comb described in section 3). The simulated FP cavity is representative of cavities produced by our group, providing ~38dB suppression at 250MHz from the central peak, and transmitting 1:120 lines (30GHz comb spacing). (right) Apparent RV bias induced by cavity transmission peak drift (detune) as measured by a resolution ($\lambda/\Delta\lambda$) 55,000 spectrograph. Since the comb lines are much narrower that the FP transmission envelope, apparent offset is dominated by asymmetry in the unresolved left and right nearest neighbors, resulting in an RV error that grows much more slowly than would be the case if the FP cavity were illuminated by a continuum source (illustrated by the near vertical dotted line).

## 2. PROGRESS TO DATE AND MOTIVATION FOR Y+J BAND INSTRUMENT

While no astronomical spectrographs currently employ frequency combs as the primary wavelength standard, numerous on-telescope demonstrations have been performed.[27,28,29,30,31] In 2010 we carried out the first astronomical RV spectroscopic observations directly supported by a laser frequency comb using an Er:fiber based comb to calibrate the Penn State Pathfinder spectrograph at the Hobby Eberly Telescope (fig. 3, 4).[14,30] The Pathfinder spectrograph was configured to provide partial coverage of the H band, extending from 1535 to 1635 nm at $\lambda/\Delta\lambda$=50,000 and the LFC covered 1450 to 1650 nm at a line spacing of 25GHz. Over the course of 6 nights 648 five minute spectra were obtained of HD16873, Sigma Draconis, Eta Cassiopeae, and Upsilon Andromeda, returning ~10m/s RV precision. The comb operated without interruption or intervention during these observations, although some adjustment was required prior to each night's observing run.

The H band comb is described in Ycas (2012)[30] and is briefly reviewed here. The comb was based on a 250 MHz Er:fiber mode locked laser whose output was initially amplified and filtered to 25 GHz over a narrow (~75 nm) band using two high finesse FP cavities. The high power output was then fed into a segment of highly non-linear fiber (HNLF) yielding an output spectrum with greater than 20 μW per mode between 1.535 and 1.635 μm. Mode filtering before the final broadening stage allowed a single FP cavity to be used for the entire band pass, rather than using two or more filters in parallel after broadening.[26]

Modal noise represented a significant source of RV uncertainty in the Pathfinder+LFC observations. While modal noise is commonly encountered in fiber fed astronomical instruments and can be controlled through the use of various approaches such as fiber agitation [32], the problem is exacerbated when feeding coherent NIR light from a small core, single mode fiber into the 300 μm core diameter multimode fibers used to feed light to Pathfinder due to interference effects and to the limited number of modes initially excited. Even with the use of a NIR optimized integrating sphere and mechanical agitation of the multimode fiber (fig. 3) measurable structure persisted that must be addressed in future instruments. Detailed study of modal noise control is currently being pursued at the University of Colorado and at the Penn State University using a variety of scrambling techniques.

The waveband selected for this initial trial (H band) enabled the rapid development of a robust, stable laser frequency comb. However, the H band is not optimal for the ultimate goal of performing RV spectroscopy on mid to late M dwarf objects. The reason for this is two fold. First, mid to late M dwarf spectra peak between 1000 and 1300 nm, with significant RV information in the J and especially Y bands. Second, the J and, again especially the Y bands display limited telluric absorption and reduced OH emission when compared to the H band.[10] For these reasons, after returning the LFC to NIST in Boulder, CO, efforts were undertaken to create a broad band comb centered at 1100 nm.

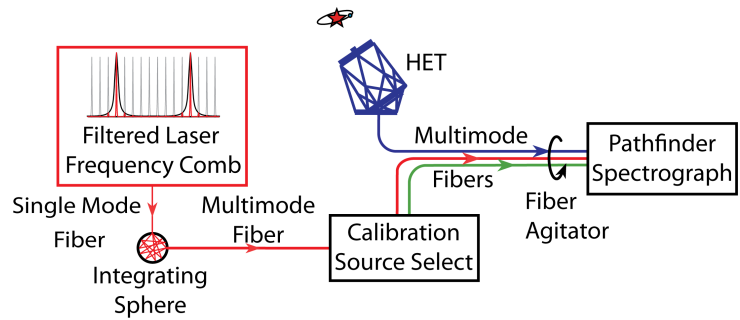

Figure 3: The 25 GHz output from the Er:fiber LFC was first fed into an integrating sphere and then coupled into multimode fiber. The frequency comb signal then passed through a free space calibration source select bench allowing the use of (and comparison with) multiple sources including a Th-Ar lamp and a U-Ne lamp.[11] From there light was transmitted to the Pathfinder spectrograph. The science and calibration fibers were offset from one another in the cross dispersion direction, allowing for simultaneous, multi-source calibration and science integration. The combination of the integrating sphere and fiber agitator was employed to reduce modal noise in the LFC spectra.

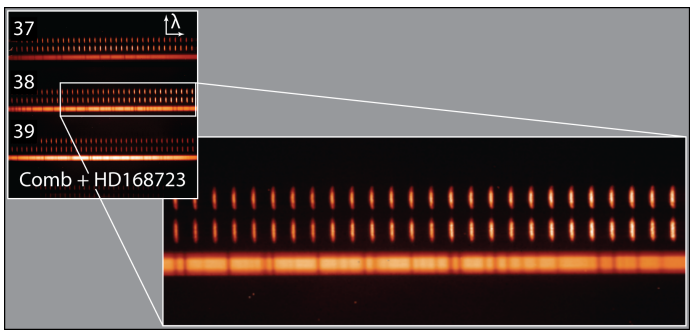

Figure 4: Sample spectrum showing both the LFC spectra (top two spectra in each order) and the stellar spectra from HD168723. Spectroscopic orders are indicated. See Ycas (2012).[30]

## 3. ERBIUM / YTTERBIUM HYBRID COMB

While in principle it is possible to broaden the output of the Er:fiber comb into the Y and J band after filtering as was done for the H band comb described in section 2, this becomes progressively more difficult at increased filter ratios and at increasing spectral offset from the center of the Er:fiber band (1550 nm). This is due to the much higher energy density required for increasingly aggressive broadening and due to the reemergence of previously suppressed modes through nonlinear processes such as four-wave mixing with increased amplification.

In the current configuration, the output from the 250 MHz Er:fiber laser is amplified and nonlinearly broadened to below 1000 nm. This is used to seed a core pumped Yb:fiber amplifier, the output of which is compressed and then coupled into a micro structured fiber resulting in a 250 MHz comb spectrum extending from 650 to 2000 nm as described in Ycas, et al.[33]

Light produced by the Rb stabilized Er:fiber laser is amplified in a core-pumped Erbium fiber amplifier to create an octave spanning spectrum with an average power of 450 mW with 70 fs pulse duration (fig. 5, red). The output is fed into a 5 cm segment of solid-core highly-nonlinear fiber (HNLF), generating a supercontinuum with ~8 % of the 240 mW power coupled into the single mode fiber falling between 1000 nm and 1100 nm. (fig. 5, blue)

The HNLF output is then coupled into a core-pumped ytterbium fiber amplifier and compressed to 70 fs pulse width. The compressed output is fed into a 0.5 m microstructured nonlinear fiber with a zero-dispersion wavelength of 945 nm, resulting in an octave-spanning spectrum, from 660 nm to 1430 nm (Fig. 5, black).

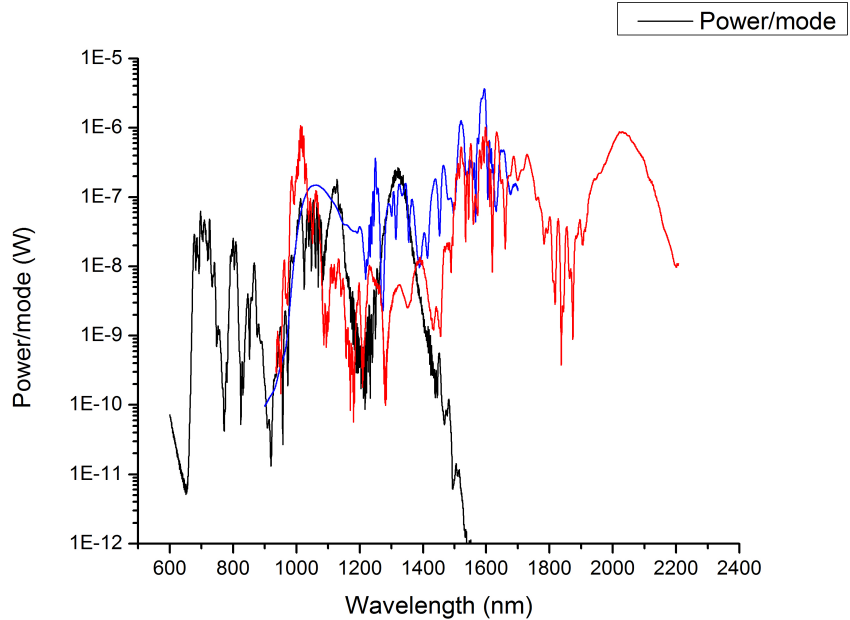

Figure 5: Erbium/Ytterbium hybrid comb output spectra. Red: Octave spanning (~950-2200 nm), broadened spectrum used for f:2f self referencing. Blue: Broadened seed spectrum for Yb amplifier (950-1700 nm). Black: Spectrum after broadening in the microstructured fiber (650-1500 nm).

## 4. FUTURE WORK

While previously we were able to filter a narrow portion of the native 1550 nm comb spectrum from 250 MHz to 25 GHz and then amplify and broaden the output to span large sections of the H band, an alternate approach is needed for the Y+J bands. Amplification of the 30 GHz comb signal to a level that a post filtered signal could be broadened is significantly less reliable for long term operations using the technology available near 1000 nm. Consequently, we will seek to amplify and broaden the spectrum at the native 250 MHz mode spacing and then operate two or more cavities in parallel[26] to filter the 250 MHz source comb spectrum to 30 GHz. Each cavity will provide finesse on the order of 300, and suppression of off-resonant modes by greater than 40 dB will be achieved by double-passing the light through each single cavity.[34] The combination of lower finesse cavities and double-passing is expected to allow us to achieve >100 nm bandwidth per filter.

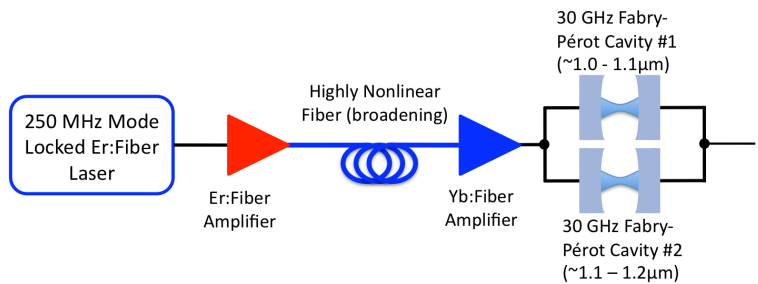

Figure 6: Output from a 250 MHz Er:fiber self referenced laser frequency comb is amplified and then broadened with no loss of coherence. The Y+J band portion of the spectrum is then amplified and will be filtered to 30GHz line spacing. Additional cavities are used to expand the filtered bandwidth.

In addition, we are pursuing several approaches to modal noise reduction: Use of non-cylindrical fibers,[35] annealed fibers,[36] large scale fiber agitation[32] and agitation of either the input or output from an integrating sphere are being explored. We anticipate that a combination of these approaches will be required to reduce modal noise to a level that does not degrade the effective precision of the frequency comb to unacceptable levels

## 5. ACKNOWLEDGEMENTS

Research presented in this paper was supported by NIST, from NASA through the NAI and Origins grant NNX09AB34G, and from the NSF grants AST-0906034, AST-1006676, AST-0907732, and AST-1126413. This work was partially supported by funding from the Center for Exoplanets and Habitable Worlds, which in turn is supported by the Pennsylvania State University, the Eberly College of Science, and the Pennsylvania Space Grant Consortium.